\documentclass{optica-article}

\journal{opticajournal} 

\articletype{Research Article}

\usepackage{lineno}
\linenumbers 

\begin{document}
\nolinenumbers
\title{One-dimensional spin-flipping topological edge state laser}

\author{Jhih-Sheng Wu\authormark{\&}, Zhen-Ting Huang, Meng-Ting Han, Yen-Hsun Chen  and Tien-Chang Lu
\authormark{*}}

\address{Department of Photonics, College of Electrical and Computer Engineering, National Yang Ming Chiao Tung University, 
No. 1001 University Road, East Dist., Hsinchu 30010, Taiwan
}

\email{\authormark{\&}jwu@nycu.edu.tw}
\email{\authormark{*}timtclu@nycu.edu.tw}


\begin{abstract*} 
Topological edge states manifest spin-momentum-locking propagation as a primary consequence of topological crystals. However, experimental studies on spin manipulation and the resulting propagation of these states are lacking. Here, we demonstrate experimentally spin manipulation of topological edge states by the boundary conditions of the one-dimensional path.  Armchair boundaries at the endpoints of the path induce spin-flipping back-scattering, resulting in a novel one-dimensional resonance --- traveling resonance. Remarkably, we demonstrate lasing of this one-dimensional traveling resonance. Our findings hold significant potential for practical applications in spin manipulation of light.
\end{abstract*}

\section{Introduction}
Topological photonics has emerged as a promising field for manipulating light propagation and exploring novel optical phenomena. The concept of topological states originates from condensed matter physics and has been extended to the realm of photonics\cite{su1979solitons,thouless1982quantized,raghu2008analogs,wang2009observation,lu2014topological,khanikaev2017two}. In recent years, significant progress has been made in realizing topological photonics and their unique characteristics in various photonic structures, such as photonic crystals and metamaterials \cite{lu2016symmetry,kim2019extremely,ota2019photonic,hafezi2013imaging,huang2022pattern}, plasmonics \cite{wu2020topological,ling2015topological,song2021plasmonic}, and lasers \cite{harari2018topological,bandres2018topological,ota2018topological,wu2020topological,zeng2020electrically,yang2022topological}. One of the unique features of topological photonics is the presence of one-dimensional (1D) topological edge states at the boundary of topological materials.  The propagation of such states demonstrates spin--momentum locking and resilience against defects and disorder \cite{hasan2010colloquium,ando2013topological,skirlo2014multimode,yang2018visualization,ozawa2019topological}, making them highly desirable for applications in information processing and communication \cite{khanikaev2013photonic,cheng2016robust,shalaev2019robust,yang2020terahertz,kumar2022phototunable}, sensing \cite{sakotic2021topological,kumar2022topological,budich2020non,koch2022quantum},  and integrated photonic devices \cite{ma2019topological,kumar2022topocom,kumar2022terahertz,yin2016realizing}.

The all-dielectrics method developed by Wu and Hu enables the implementation of topological photonics across a wide range of frequencies \cite{wu2015scheme}. It has been demonstrated that at optical frequencies, topological edge states and spin-momentum locking are achievable\cite{saby2018,Nik20}. These studies have confirmed the robust connections between pseudospins, propagation directions, and far-field polarization through experimental demonstrations. Specifically, states of a particular spin can only propagate in a one-way direction. Such one-way transports could have great potential for chiral quantum optics and information processing \cite{lodahl2017chiral,yang2020spin}. However, the process of switching spins in topological photonics requires further investigation. Since the crystal symmetry protects the spin--momentum locking, spin--flipping is possible by breaking the crystal symmetries. This work investigates strong spin-flipping and back-scattering due to breaking the $C_6$ symmetry at the endpoints of 1D topological edge states. With this strong spin-flipping and back-scattering, we design a 1D topological cavity that can produce laser light.

We introduce a novel type of topological cavity laser that utilizes spin-flipping reflection at the ends of a 1D cavity. Instead of a closed path, we employ an open-end 1D path for the topological edge states. The endpoints of the 1D cavity break the $C_6$ symmetry, leading to the reflection and spin-flipping of the topological edge states.  Traveling and spin-flipping reflections generate a new type of resonance called ``traveling wave resonance.'' We demonstrate the lasing characteristics of this traveling wave resonance.

\section{Schematics and Methods}
\subsection{Schematics and physical pictures}
\begin{figure}[htbp]
\centering\includegraphics[width=8cm]{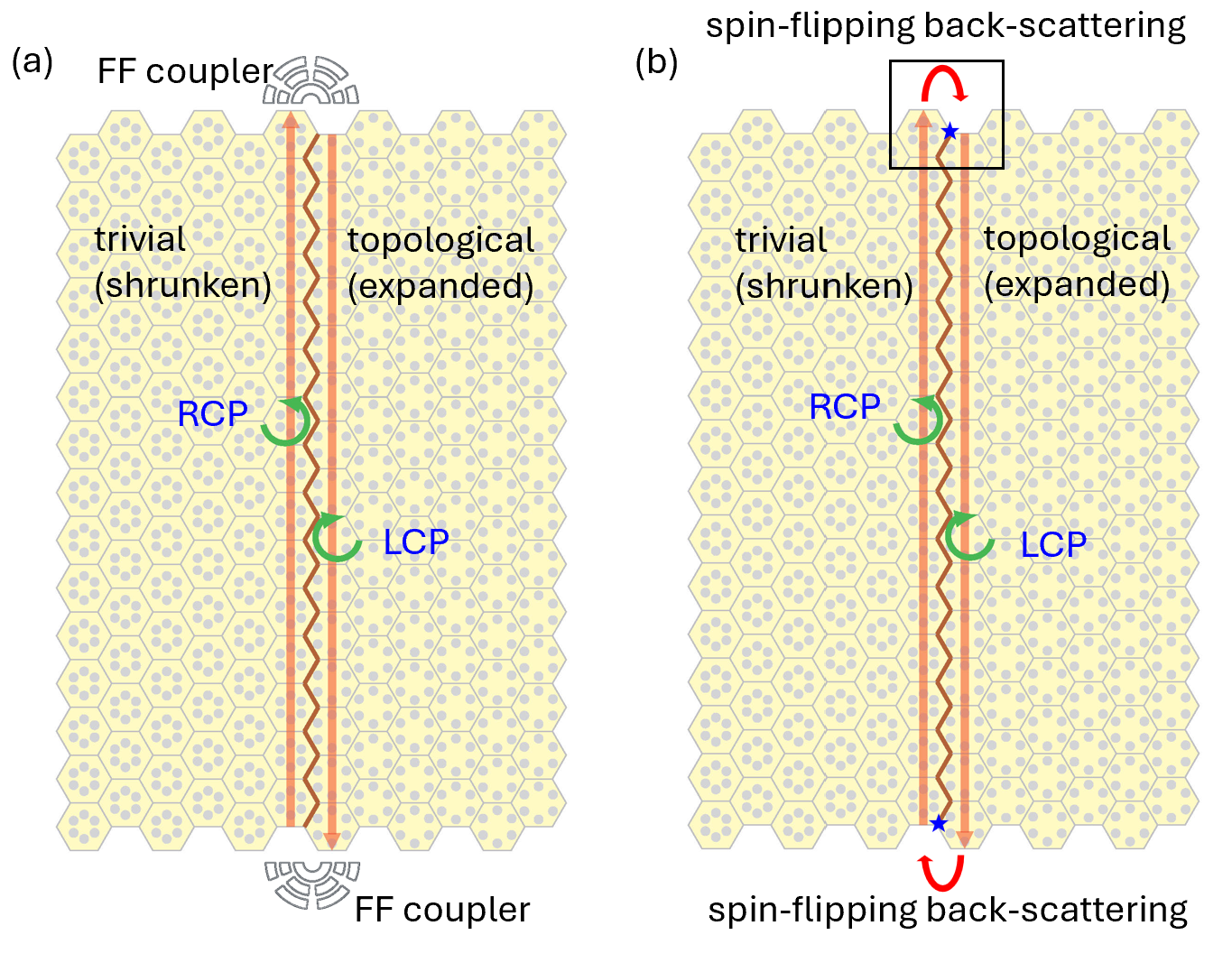}
\caption{Topological edge states (TESs) with and without spin-flipping back-scattering. (a) Topological edge states form at the interface of the trivial (expanded) and topological (shrunken) lattices. The zigzag brown line denotes the interface. With the far-field (FF) couplers, TESs are converted to LCP and RCP polarized emissions at the top and bottom, respectively. Such emissions confirm the spin-momenting locking property. (b) The dielectric structure near the endpoints (the blue stars), breaking $C_6$ symmetry, can efficiently reflect TESs and flip their spins. A closed round trip of TESs results in a 1D traveling wave resonance. We have shown that this type of resonance can lead to laser action.}\label{fig:sch}
\end{figure}

Topological (trivial) photonic bands can be formed by expanding (shrinking) the hexagons of the triangular lattices.  Figure~\ref{fig:sch}(a) shows topological edge states (TESs) at the interface of the trivial and topological lattices. The TESs, with the spin-momentum locking property, are out-coupled to the left-circular polarization (LCP) and right-circular polarization (RCP) far-field (FF) radiations, respectively. Up-going TESs with the RCPs are incident on the top FF coupler and converted to the RCP FF radiations, while down-going TESs with the LCPs are incident on the bottom FF coupler and converted to the LCP FF radiations. To have spin-flipping back-scattering, the structures near the endpoints of TESs must be designed appropriately. 
First, We consider a structure without the FF coupler in Fig. \ref{fig:sch}(b).  The structures near the endpoints (the blue stars in Fig. \ref{fig:sch}(b)) are chosen intentionally to break the $C_6$ symmetry to reflect TESs and flip their spins efficiently. Theoretically, breaking the $C_6$ symmetry means that pseudo-spins of the up-going and down-going TESs are mixed, and thus back-scattering is allowed. It is important to note that the reflected TESs, possessing the opposite spin, cannot interfere with the incident TESs. Consequently, rather than generating a standing wave resonance, TESs that travel and reflect twice follow a closed path and establish a traveling wave resonance. This phenomenon is similar to the resonances observed in a laser gyroscope. We refer to this phenomenon as ``1D spin-flipping traveling wave resonance."  

The up-going and down-going TESs, $|k_u,-\rangle$ and $|k_d,+\rangle$, are the superpositions of the circular dipole and quadrupole states. They can be expressed as \cite{wu2015scheme}
\begin{align}
|k_u,+\rangle &= (\alpha |p_+\rangle +\beta |d_+\rangle)e^{ik_u x},\\
|k_d,-\rangle &= (\alpha^{*} |p_-\rangle +\beta^{*} |d_-\rangle)e^{ik_d x},
\end{align}
where $k_u =-k_d$ is the wave vector, $\pm$ denotes the pseudospins, and $p_\pm$ and $d_\pm$ are the circular dipole and quadrupole modes, as shown in Fig.~\ref{fig:topo_band}. Due to the pseudo-time-reversal symmetry, back-scattering does not occur if the structure maintains this symmetry. To induce back-scattering, one can break the pseudo-spin symmetry ($C_6$) by introducing a suitable $\epsilon(\mathbf{r})$ such that $\langle p_+|\epsilon(\mathbf{r})|p_-\rangle$ is not equal to zero. Given that $|p_\pm\rangle\sim e^{\pm i\theta}$, the dielectric function $\epsilon(\mathbf{r})$ should contain the $e^{\pm i2\theta}$ components in order for $\langle p_+|\epsilon(\mathbf{r})|p_-\rangle$ to be non-zero. In Fig. \ref{fig:sch}(b), the TESs encounter scattering at the endpoints (blue stars). By using the blue stars as the centers in Fig.~\ref{fig:sch}(b), one can compute the angular decomposition of $\epsilon(\mathbf{r})$
\begin{align}
    a_n = \int d^2r \epsilon(\mathbf{r}) e^{in\theta}  
\end{align}
and show that the $e^{\pm i2\theta}$ components of the dielectric function $\epsilon(\mathbf{r})$ are non-zero. Therefore, spin-flipping can occur due to the change of the circular polarization by $\langle p_+|\epsilon(\mathbf{r})|p_-\rangle$.
Because spin-flipping of the TESs is accompanied by a change of momentum, spin-flipping back-scattering occurs due to the endpoints in Fig.~\ref{fig:sch}(b).   

\subsection{Topological crystal design }
\begin{figure}[htbp]
\centering\includegraphics[width=12cm]{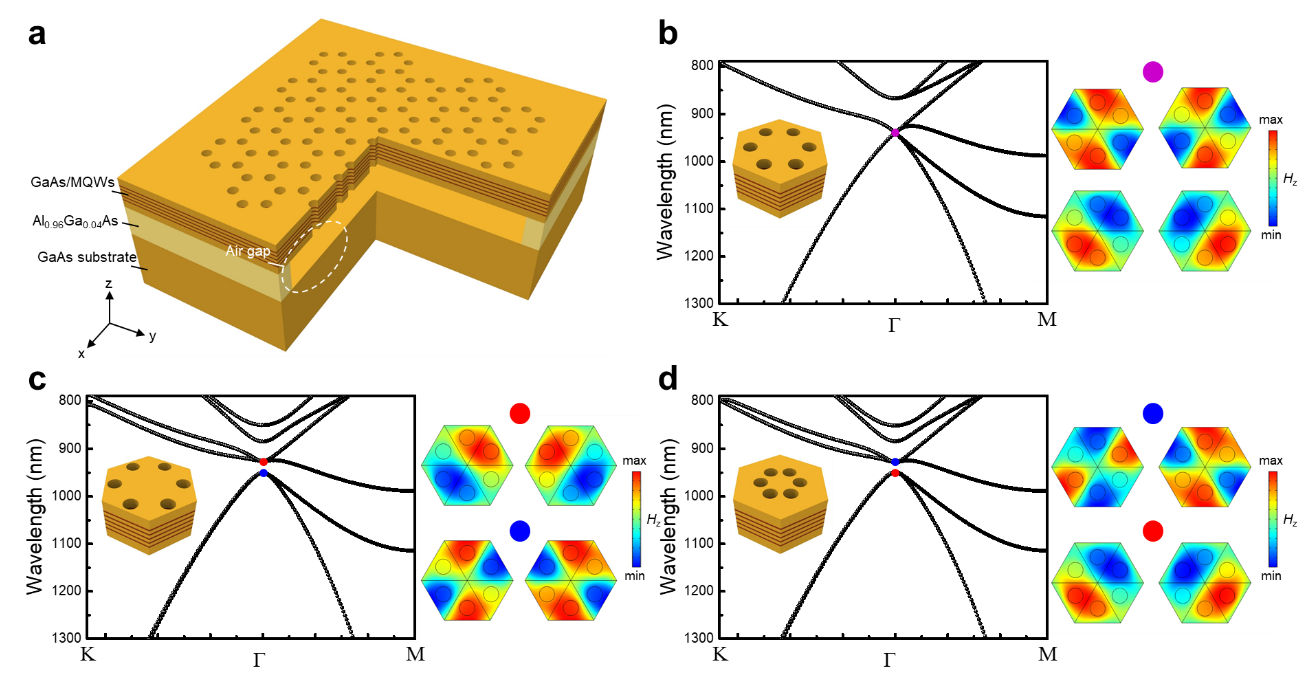}
\caption{Schematics and band structures of the trivial and topological lattice. (a) Schematic diagram of the suspended GaAs/MQWs membrane with air holes arranged in the honeycomb lattice. (b) Band structure of the honeycomb lattice, where the doubly-degenerate Dirac cones can be observed at the $\Gamma$ point. The right insets illustrate the out-of-plane magnetic field ($H_z$) distribution of the degenerate dipole (lower row) and quadrupole modes (upper row) at the Dirac point. (c),(d) Band structures of the expanded (c) and shrunken (d) honeycomb lattices. These two deformed lattices open a band gap at the Dirac point and separate the dipole and quadrupole modes. The one that makes the quadrupole modes stay in the lower energy level belongs to the topological lattice (c), and the other belongs to the trivial lattice (d).}\label{fig:topo_band}
\end{figure}

The schematics of our topological photonic crystals (PCs) and experimental fabrication are shown in Fig.~\ref{fig:topo_band}.  
Figure \ref{fig:topo_band}(a) illustrates the suspended GaAs membrane schematic diagram with five pairs of InGaAs/GaAs multiple quantum wells (MQWs). The air gap etched in the sacrificial AlGaAs layer provides a significant enough refractive index difference between the membrane and air to confine the optical field well. The photoluminescence (PL) spectrum of the MQWs peaked at 925~nm, as shown in Fig. S1 of Supplement~1.
On the suspended membrane, air holes were fabricated and arranged in the honeycomb lattice, whose primitive unit cell is a rhombus and consists of two air holes. 
Based on this unit cell, Dirac cones can be observed at K and K$^\prime$ valleys in the reciprocal space.
We design the materials and geometries such that the Dirac points are at $\lambda = 938$~nm. The lattice constant and filling factor of the expanded honeycomb lattice are 383~nm and 0.2, respectively. 

Topological and trivial phases of matter can be achieved by expanding and shrinking a hexagonal unit of the honeycomb lattice \cite{wu2015scheme}. Without expansion and shrinking, Dirac cones at the K and K$^\prime$ valleys in the Brillouin zone, as defined by a rhombic unit cell, are located at the $\Gamma$ point in the shrunken Brillouin zone defined by a hexagonal unit cell. 
Figure~\ref{fig:topo_band}(b) shows the band structure of the hexagonal unit cell, illustrating the presence of doubly degenerate Dirac cones at the $\Gamma$ point. There are four degenerate modes at the Dirac point: two belong to the dipole modes $|p_\pm\rangle$, and the others belong to the quadrupole modes $|d_\pm\rangle$. The mode profiles are shown in the small images on the right of Figure~\ref{fig:topo_band}(b). When the lattice expands or shrinks, the air holes move 5~nm from the center of the hexagon. Expanding or shrinking the lattice to open a band gap at the Dirac point can make the photonic crystal (PC) either topological or trivial. Figures~\ref{fig:topo_band}(c) and (d) display the band structures of the expanded and shrunken honeycomb lattices, respectively. In the topological lattice (Figure~\ref{fig:topo_band}(c)), dipole modes are above the band gap, while the quadrupole modes are below. In the trivial lattice (Figure~\ref{fig:topo_band}(d)), the sequence is reversed, indicating band inversion and a change in topology.

\begin{figure}[htbp]
\centering\includegraphics[width=10cm]{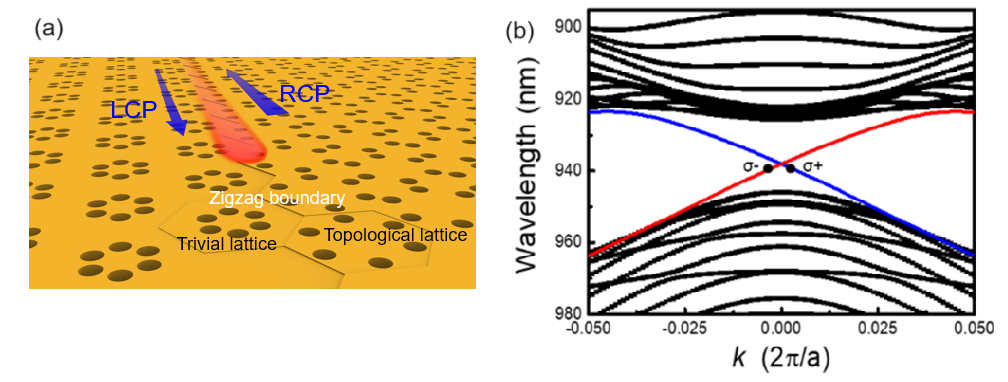}
\caption{Dispersions of the TESs, with spin-momentum locking. $\sigma_\pm$ denotes RCP and LCP, respectively. (a) The interface of the topological and trivial lattices hosts the spin-momentum-locked TESs.  Up-going (positive $k$ direction) TESs have the RCP ($\sigma_+$). (b)
The red (blue) line shows the dispersion of the TES with a $\sigma_+$ spin ($\sigma_-$ spin). 
TESs with a $\sigma_+$ spin have a positive group velocity, while 
TESs with a $\sigma_-$ spin have a negative group velocity. The black lines are the dispersions of the bulk bands.
}\label{fig:TES_band}
\end{figure}

Topological edge states (TESs) emerge at the interface between topological insulators and trivial insulators. As shown in Figure~\ref{fig:TES_band}, the depicted band structures elucidate the distinction between the bulk states and TESs. TESs manifest within the bandgap, featuring a crossing at the $\Gamma$ point. This intersection is preserved by pseudo-time-reversal symmetry. Furthermore, there is a direct correlation between the spin orientations and the propagation directions (group velocities) of these states, indicating the spin-momentum locking.

\subsection{Fabrication} 
We used the metal-organic chemical vapor deposition (MOCVD) system to grow the epitaxial structure on a 2" GaAs wafer. The epitaxial growth started with a buffer GaAs layer, followed by a 2~{\textmu}m thick Al$_{0.96}$Ga$_{0.04}$As sacrificial layer. The 200~nm thick GaAs layer contains five pairs of InGaAs/GaAs MQWs. The thickness of InGaAs QW and GaAs barrier were 6~nm and 10~nm, respectively. InGaAs QW and GaAs barrier thickness  The honeycomb lattice was fabricated on the as-grown wafer through electron beam lithography and inductively coupled plasma dry etching. The etching depth was around 1.2~{\textmu}m in order to expose the underneath sacrificial layer. The suspended membrane was fabricated through the wet etching from those lattice holes and etching assistance via holes around the honeycomb lattice patterns, where a sacrificial layer of Al$_{0.96}$Ga$_{0.04}$As was etched by immersing the sample into the buffered oxide etchant (BOE) for 3 minutes, which was mixed with hydrofluoric acid (HF) and ammonium fluoride (NH4F) at a mixing ratio of 1:6. The sample was then placed still until it was dry out and then sent for optical measurement.

\subsection{Measurement}
Optical pumping was performed using a 532~nm pulse laser with a repetition rate of 1 kHz and a pulse duration of 0.35~ns at room temperature. 
To measure the far-field spectra maps, a 4f system was introduced into our micro-photoluminescence (\textmu-PL) setup. A 4f system was introduced into our micro-photoluminescence (\textmu-PL) setup to measure the far-field spectra maps. When the laser beam is focused on the sample surface, the emission will be imaged at a back focal plane behind the objective lens. The emission will be imaged behind the objective lens at a back focal plane when the laser beam is focused on the sample surface. To further image the back focal plane to the charge-coupled device (CCD), two extra lenses with the same focal length were added in front of the spectrometer. The configuration of lenses forms the 4f system, and then the back focal plane can be imaged to the entrance slit of the spectrometer based on the Fourier optics. Afterward, the back focal plane can be imaged to the CCD, and the angle-resolved far-field spectra maps can be observed. The angle resolution and spectral resolution were estimated to be 1$^\circ$ and 0.1~nm, respectively.

\section{ Results and discussions}

We initially showcased the spin-momentum-locked TESs in our device interface as depicted in Fig.~\ref{fig:sch}(a). Subsequently, we presented a 1D spin-flipping TES laser attributed to spin-flipping back-scattering, shown schematically in Fig.~\ref{fig:sch}(b).
\begin{figure}[htbp]
\centering\includegraphics[width=14cm]{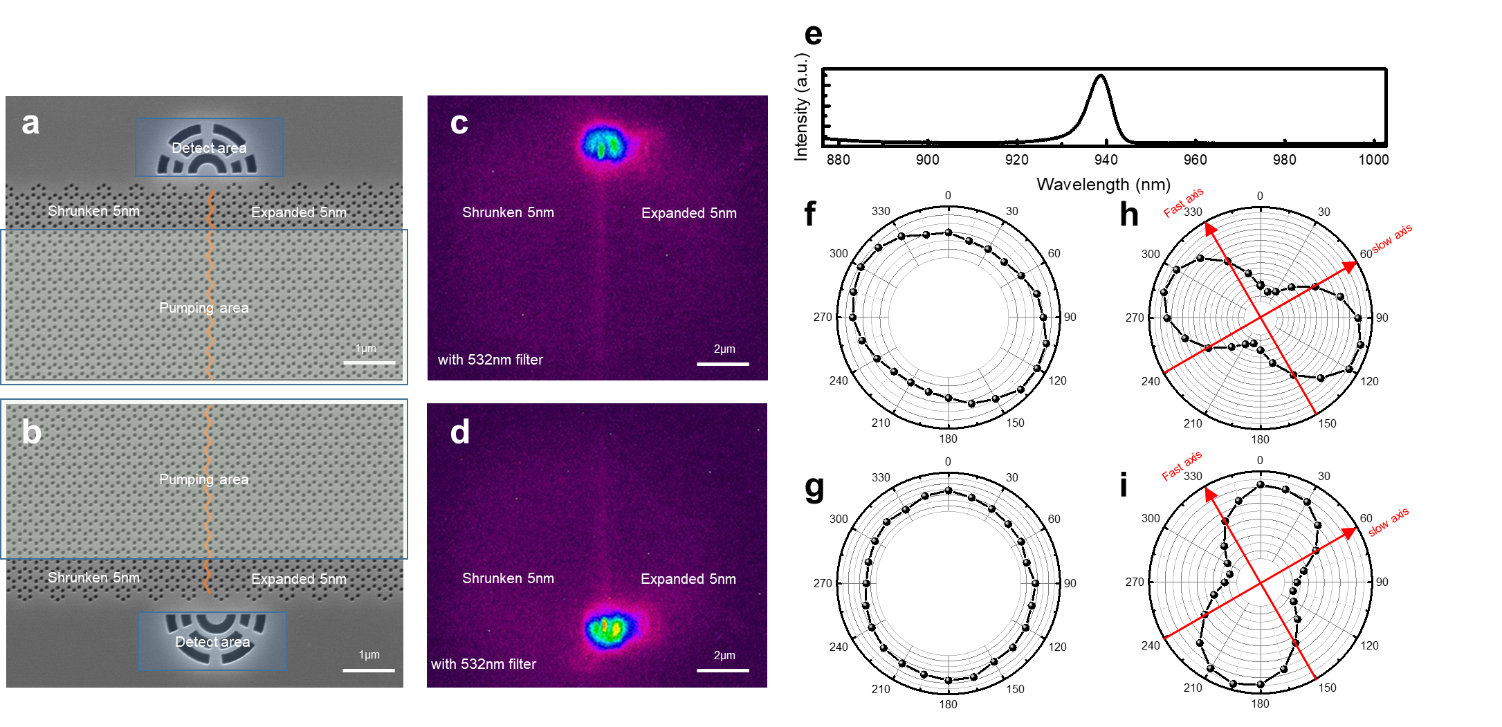}
\caption{Spin-momentum-locking of traveling waves. (a) and (b) are the SEM images of the top and bottom FF couplers, which convert the TES to the LCP/RCP emissions. (c) and (d) are the near-field images of (a) and (b), respectively. (e) PL spectrum of the top FF coupler. (f) and (g): polarizations of the top and bottom couplers, respectively. (h) and (i): polarizations of (f) and (g) after a QWP.}\label{fig:sml}
\end{figure}

\subsection{Spin-momentum locking of traveling waves}
Figure~\ref{fig:sml}(a) and (b) illustrate the SEM images of the combined topological and trivial crystals, of which size is 100~{\textmu}m $\times$ 100~{\textmu}m. To measure the spin-momentum-locked TESs, two grating structures were fabricated at the upper and lower ends of the combined topological and trivial crystals \cite{saby2018}. These grating structures act as output couplers, allowing the topological edge states to be coupled into the radiation and detected by the objective. Subsequently, we optically pumped the combined topological and trivial crystals in the central region and detected the light at the upper or lower output grating coupler. The diameter of the pumping spot was around 80~{\textmu}m. Figures~\ref{fig:sml}(c) and (d) illustrate the measured near-field images after passing through a 532~nm long-pass filter. Remarkably, in addition to the bright spots in the grating regions, a distinct bright line can also be observed along the interface, resulting from the propagation of the TESs. The PL spectrum measured from the upper output grating coupler is illustrated in Fig.~\ref{fig:sml}(e), where the PL peak is 939 nm. Figures~\ref{fig:sml}(f) and (g) illustrate the FF polarization measured from the upper and lower output grating couplers, respectively, which all show circular polarizations.
A quarter-wave plate (QWP) was placed in the optical path to determine the LCP or RCP.
After passing through a QWP, the FF polarization in Fig.~\ref{fig:sml}(f) becomes linearly polarized at -45$^\circ$ to the fast axis, and that in Fig.~\ref{fig:sml}(g) becomes linearly polarized at +45$^\circ$ to the fast axis, as illustrated in Fig.~\ref{fig:sml}(h) and (i). These results indicate that the TESs propagating towards the upper and lower output grating couplers have opposite circular polarizations, precisely right circular polarization (RCP) and left circular polarization (LCP), respectively. The above results confirm the spin-momentum locking of TESs of our samples.

\subsection{Lasing of 1D traveling wave topological edge
state}

\begin{figure}[htbp]
\centering\includegraphics[width=13.5cm]{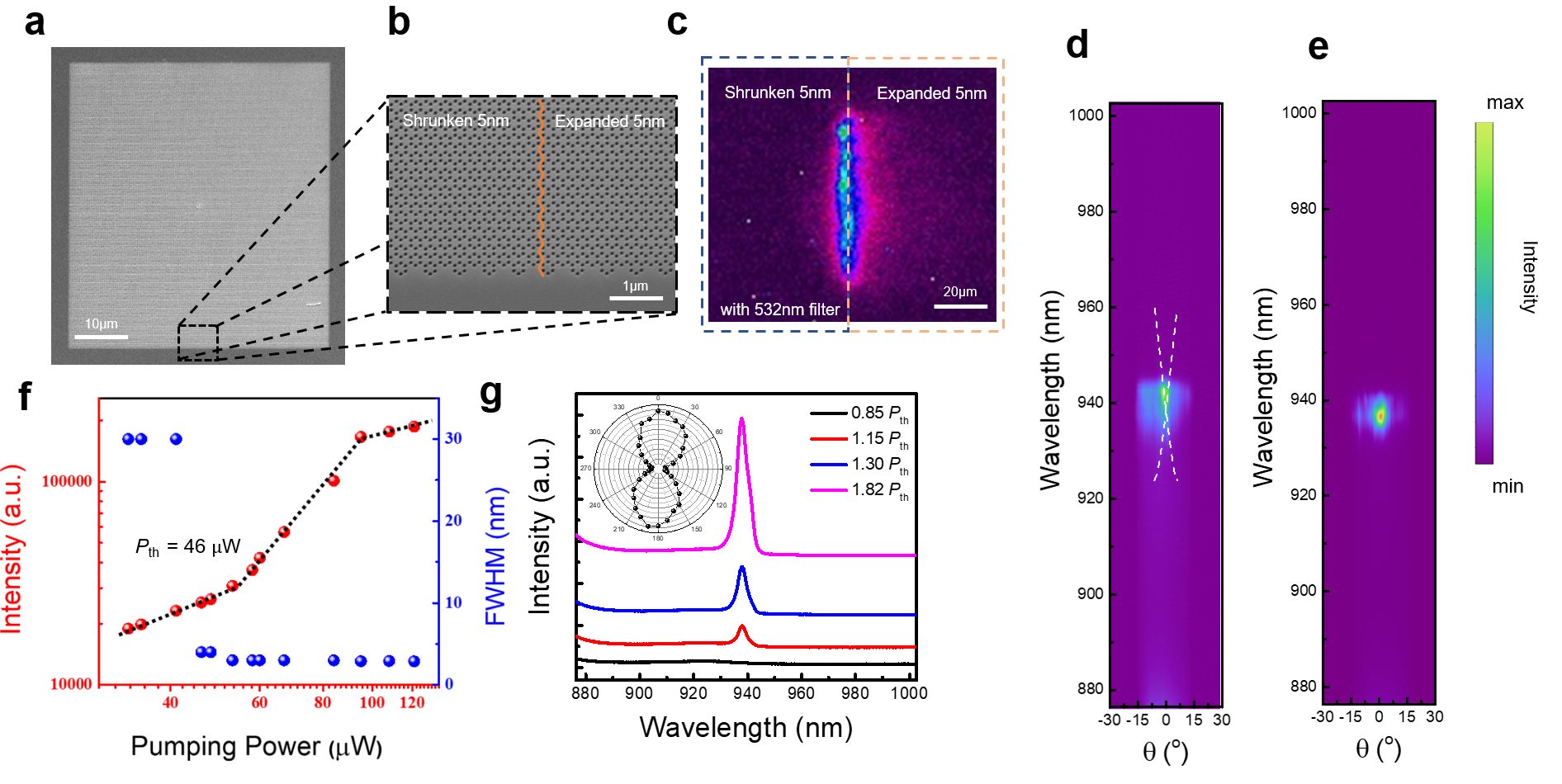}
\caption{Characteristics of the 1D spin-flipping toplogical laser. (a),(b) SEM images of the combined topological and trivial crystals, composed of shrunken and expanded honeycomb lattices. The endpoints violate the $C_6$ symmetry, enabling the spin-flipping back-scattering. (c) Near-field of the TES lasing. (d),(e) FF spectra maps below (d) and above (e) the pumping threshold. The dashed line in (d) indicates the dispersion of topological edge states. (f) L-L curve and power-dependent PL spectra (g) of the combined topological and trivial crystals, where the pumping threshold is around 46~{\textmu}W. The inset in (g) indicates the far-field polarization.} \label{fig:1dtw_laser}
\end{figure}
Spin-flipping back-scattering is allowed if the structure violates the $C_6$ symmetry \cite{Nik20}.
To have the proposed 1D traveling wave resonance, we fabricated the combined topological and trivial crystals without an output grating coupler, as shown in Figs.~\ref{fig:1dtw_laser}(a) and (b). The structure near the endpoint (Figs.~\ref{fig:1dtw_laser}(b)) violates the $C_6$ symmetry and enables back-scattering. A round trip occurs due to the top and bottom back-scattering. This round-trip oscillation does not support the standing wave pattern because the up-going and down-going TESs have opposite pseudospins. Instead, 
a resonance of the traveling wave is constructed, analogous to laser gyroscopes. Figures~\ref{fig:1dtw_laser}(a) and (b) illustrate the SEM images of our samples, of which size is 100~{\textmu}m~$\times$~100~{\textmu}m. Then, we performed the optical pumping and detected the light in the central region at the interface. Figure~\ref{fig:1dtw_laser}(c) illustrates the near-field images after passing through a 532~nm long-pass filter. A clear, bright line can be observed at the interface, resulting from the traveling-wave resonance of the TESs. Figure~\ref{fig:1dtw_laser}(d) and (e) illustrate the FF spectra maps below and above the pumping threshold. Below the pumping threshold, the FF spectra map has the crossing, which is the dispersions of TESs as predicted in Fig.~\ref{fig:TES_band}(b). When the pumping power is above the threshold ($P_\textrm{th}=46$~{\textmu}W), a strong spot appears at 939 nm and 0$^\circ$ in Fig.~\ref{fig:TES_band}(e). We conducted measurements of the light-in versus light-out (L-L) curve and the full width at half maximum (FWHM) of the PL spectra to determine whether a specific spot results from lasing action. In Fi~g.\ref{fig:1dtw_laser}(f), the L-L curve and the FWHMs show evidence of lasing action of the 1D TESs. The pumping threshold is 46~{\textmu}W, and the lasing linewidth is 3~nm. In the inset of Fig.~\ref{fig:1dtw_laser}(g), the FF polarization is presented without passing through a QWP. This demonstrates the linear polarization resulting from the combination of left-handed circularly polarized (LCP) and right-handed circularly polarized (RCP) light emission. The linear polarization at the $\Gamma$ point is evidence of 1D spin-flipping traveling wave resonance.

These findings lay the groundwork for manipulating and converting the topological pseudospins. 

\section{Conclusions}
We have developed and demonstrated a new type of 1D TES laser. This laser's resonance is created by combining spin-flipping back-scattering and traveling waves. We have fabricated a photonic crystal with a 1D interface between a topological and a trivial lattice. The endpoints of this 1D interface play a crucial role in spin-flipping back-scattering. We induce back-scattering by intentionally breaking the $C_6$ symmetry of our structure. The round trip of the traveling waves forms a resonance. We have experimentally demonstrated the lasing of this resonance, which exhibits linear polarization. This confirms that the lasing produces both LCP and RCP emissions, indicating spin-flipping. Our work is significant for understanding the manipulation of TESs and their spins. These findings will have implications for information processing using topological photonics.

\newpage

 \begin{backmatter}
 \bmsection{Funding}
 This work was financially supported by the National Science and Technology Council of Taiwan
under Contract Nos. NSTC 113-2221-E-A49 -067 -MY3 and MOST 111-2112-M-A49 -015 -MY3.
 \end{backmatter}
\section{Supplement 1}

\setcounter{figure}{0}

\makeatletter
\renewcommand{\thefigure}{S\@arabic\c@figure}
\makeatother
\begin{figure}[htbp]
\centering\includegraphics[width=10cm]{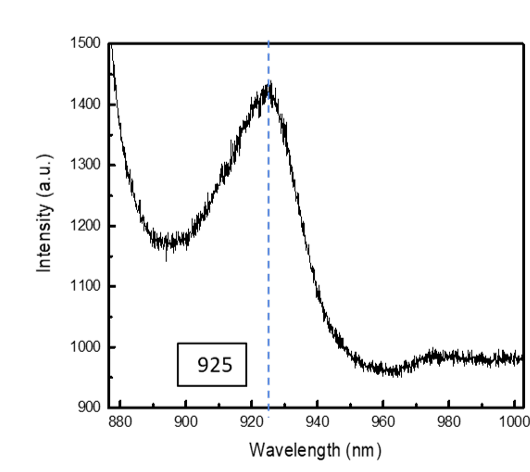}
\caption{The photoluminescence (PL) spectrum of the suspended GaAs membrane with five pairs of InGaAs/GaAs multiple quantum wells (MQWs), where the PL peak is located at 925 nm.}\label{fig:pl}
\end{figure}

\begin{thebibliography}{10}
\newcommand{\enquote}[1]{``#1''}

\bibitem{su1979solitons}
W.-P. Su, J.~R. Schrieffer, and A.~J. Heeger, \enquote{Solitons in
  polyacetylene,} {\protect\JournalTitle{Physical Review Letters}} \textbf{42},
  1698 (1979).

\bibitem{thouless1982quantized}
D.~J. Thouless, M.~Kohmoto, M.~P. Nightingale, and M.~den Nijs,
  \enquote{Quantized hall conductance in a two-dimensional periodic potential,}
  {\protect\JournalTitle{Physical Review Letters}} \textbf{49}, 405 (1982).

\bibitem{raghu2008analogs}
S.~Raghu and F.~D.~M. Haldane, \enquote{Analogs of quantum-hall-effect edge
  states in photonic crystals,} {\protect\JournalTitle{Physical Review A}}
  \textbf{78}, 033834 (2008).

\bibitem{wang2009observation}
Z.~Wang, Y.~Chong, J.~D. Joannopoulos, and M.~Solja{\v{c}}i{\'c},
  \enquote{Observation of unidirectional backscattering-immune topological
  electromagnetic states,} {\protect\JournalTitle{Nature}} \textbf{461},
  772--775 (2009).

\bibitem{lu2014topological}
L.~Lu, J.~D. Joannopoulos, and M.~Solja{\v{c}}i{\'c}, \enquote{Topological
  photonics,} {\protect\JournalTitle{Nature Photonics}} \textbf{8}, 821--829
  (2014).

\bibitem{khanikaev2017two}
A.~B. Khanikaev and G.~Shvets, \enquote{Two-dimensional topological photonics,}
  {\protect\JournalTitle{Nature Photonics}} \textbf{11}, 763--773 (2017).

\bibitem{lu2016symmetry}
L.~Lu, C.~Fang, L.~Fu, \emph{et~al.}, \enquote{Symmetry-protected topological
  photonic crystal in three dimensions,} {\protect\JournalTitle{Nature
  Physics}} \textbf{12}, 337--340 (2016).

\bibitem{kim2019extremely}
M.~Kim, W.~Gao, D.~Lee, \emph{et~al.}, \enquote{Extremely broadband topological
  surface states in a photonic topological metamaterial,}
  {\protect\JournalTitle{Advanced Optical Materials}} \textbf{7}, 1900900
  (2019).

\bibitem{ota2019photonic}
Y.~Ota, F.~Liu, R.~Katsumi, \emph{et~al.}, \enquote{Photonic crystal nanocavity
  based on a topological corner state,} {\protect\JournalTitle{Optica}}
  \textbf{6}, 786--789 (2019).

\bibitem{hafezi2013imaging}
M.~Hafezi, S.~Mittal, J.~Fan, \emph{et~al.}, \enquote{Imaging topological edge
  states in silicon photonics,} {\protect\JournalTitle{Nature Photonics}}
  \textbf{7}, 1001--1005 (2013).

\bibitem{huang2022pattern}
Z.-T. Huang, K.-B. Hong, R.-K. Lee, \emph{et~al.}, \enquote{Pattern-tunable
  synthetic gauge fields in topological photonic graphene,}
  {\protect\JournalTitle{Nanophotonics}} \textbf{11}, 1297--1308 (2022).

\bibitem{wu2020topological}
J.-S. Wu, V.~Apalkov, and M.~I. Stockman, \enquote{Topological spaser,}
  {\protect\JournalTitle{Physical Review Letters}} \textbf{124}, 017701 (2020).

\bibitem{ling2015topological}
C.~Ling, M.~Xiao, C.~T. Chan, \emph{et~al.}, \enquote{Topological edge plasmon
  modes between diatomic chains of plasmonic nanoparticles,}
  {\protect\JournalTitle{Optics Express}} \textbf{23}, 2021--2031 (2015).

\bibitem{song2021plasmonic}
Q.~Song, M.~Odeh, J.~Z{\'u}{\~n}iga-P{\'e}rez, \emph{et~al.},
  \enquote{Plasmonic topological metasurface by encircling an exceptional
  point,} {\protect\JournalTitle{Science}} \textbf{373}, 1133--1137 (2021).

\bibitem{harari2018topological}
G.~Harari, M.~A. Bandres, Y.~Lumer, \emph{et~al.}, \enquote{Topological
  insulator laser: theory,} {\protect\JournalTitle{Science}} \textbf{359},
  eaar4003 (2018).

\bibitem{bandres2018topological}
M.~A. Bandres, S.~Wittek, G.~Harari, \emph{et~al.}, \enquote{Topological
  insulator laser: Experiments,} {\protect\JournalTitle{Science}} \textbf{359},
  eaar4005 (2018).

\bibitem{ota2018topological}
Y.~Ota, R.~Katsumi, K.~Watanabe, \emph{et~al.}, \enquote{Topological photonic
  crystal nanocavity laser,} {\protect\JournalTitle{Communications Physics}}
  \textbf{1}, 86 (2018).

\bibitem{zeng2020electrically}
Y.~Zeng, U.~Chattopadhyay, B.~Zhu, \emph{et~al.}, \enquote{Electrically pumped
  topological laser with valley edge modes,} {\protect\JournalTitle{Nature}}
  \textbf{578}, 246--250 (2020).

\bibitem{yang2022topological}
L.~Yang, G.~Li, X.~Gao, and L.~Lu, \enquote{Topological-cavity surface-emitting
  laser,} {\protect\JournalTitle{Nature Photonics}} \textbf{16}, 279--283
  (2022).

\bibitem{hasan2010colloquium}
M.~Z. Hasan and C.~L. Kane, \enquote{Colloquium: topological insulators,}
  {\protect\JournalTitle{Reviews of Modern Physics}} \textbf{82}, 3045--3067
  (2010).

\bibitem{ando2013topological}
Y.~Ando, \enquote{Topological insulator materials,}
  {\protect\JournalTitle{Journal of the Physical Society of Japan}}
  \textbf{82}, 102001 (2013).

\bibitem{skirlo2014multimode}
S.~A. Skirlo, L.~Lu, and M.~Solja{\v{c}}i{\'c}, \enquote{Multimode one-way
  waveguides of large chern numbers,} {\protect\JournalTitle{Physical Review
  Letters}} \textbf{113}, 113904 (2014).

\bibitem{yang2018visualization}
Y.~Yang, Y.~F. Xu, T.~Xu, \emph{et~al.}, \enquote{Visualization of a
  unidirectional electromagnetic waveguide using topological photonic crystals
  made of dielectric materials,} {\protect\JournalTitle{Physical Review
  Letters}} \textbf{120}, 217401 (2018).

\bibitem{ozawa2019topological}
T.~Ozawa, H.~M. Price, A.~Amo, \emph{et~al.}, \enquote{Topological photonics,}
  {\protect\JournalTitle{Reviews of Modern Physics}} \textbf{91}, 015006
  (2019).

\bibitem{khanikaev2013photonic}
A.~B. Khanikaev, S.~Hossein~Mousavi, W.-K. Tse, \emph{et~al.},
  \enquote{Photonic topological insulators,} {\protect\JournalTitle{Nature
  Materials}} \textbf{12}, 233--239 (2013).

\bibitem{cheng2016robust}
X.~Cheng, C.~Jouvaud, X.~Ni, \emph{et~al.}, \enquote{Robust reconfigurable
  electromagnetic pathways within a photonic topological insulator,}
  {\protect\JournalTitle{Nature Materials}} \textbf{15}, 542--548 (2016).

\bibitem{shalaev2019robust}
M.~I. Shalaev, W.~Walasik, A.~Tsukernik, \emph{et~al.}, \enquote{Robust
  topologically protected transport in photonic crystals at telecommunication
  wavelengths,} {\protect\JournalTitle{Nature Nanotechnology}} \textbf{14},
  31--34 (2019).

\bibitem{yang2020terahertz}
Y.~Yang, Y.~Yamagami, X.~Yu, \emph{et~al.}, \enquote{Terahertz topological
  photonics for on-chip communication,} {\protect\JournalTitle{Nature
  Photonics}} \textbf{14}, 446--451 (2020).

\bibitem{kumar2022phototunable}
A.~Kumar, M.~Gupta, P.~Pitchappa, \emph{et~al.}, \enquote{Phototunable
  chip-scale topological photonics: 160 gbps waveguide and demultiplexer for
  thz 6g communication,} {\protect\JournalTitle{Nature communications}}
  \textbf{13}, 5404 (2022).

\bibitem{sakotic2021topological}
Z.~Sakotic, A.~Krasnok, A.~Al{\'u}, and N.~Jankovic, \enquote{Topological
  scattering singularities and embedded eigenstates for polarization control
  and sensing applications,} {\protect\JournalTitle{Photonics Research}}
  \textbf{9}, 1310--1323 (2021).

\bibitem{kumar2022topological}
A.~Kumar, M.~Gupta, P.~Pitchappa, \emph{et~al.}, \enquote{Topological sensor on
  a silicon chip,} {\protect\JournalTitle{Applied Physics Letters}}
  \textbf{121} (2022).

\bibitem{budich2020non}
J.~C. Budich and E.~J. Bergholtz, \enquote{Non-hermitian topological sensors,}
  {\protect\JournalTitle{Physical Review Letters}} \textbf{125}, 180403 (2020).

\bibitem{koch2022quantum}
F.~Koch and J.~C. Budich, \enquote{Quantum non-hermitian topological sensors,}
  {\protect\JournalTitle{Physical Review Research}} \textbf{4}, 013113 (2022).

\bibitem{ma2019topological}
J.~Ma, X.~Xi, and X.~Sun, \enquote{Topological photonic integrated circuits
  based on valley kink states,} {\protect\JournalTitle{Laser \& Photonics
  Reviews}} \textbf{13}, 1900087 (2019).

\bibitem{kumar2022topocom}
A.~Kumar, M.~Gupta, and R.~Singh, \enquote{Topological integrated circuits for
  5{G} and 6{G},} {\protect\JournalTitle{Nature Electronics}} \textbf{5},
  261--262 (2022).

\bibitem{kumar2022terahertz}
A.~Kumar, M.~Gupta, P.~Pitchappa, \emph{et~al.}, \enquote{Terahertz topological
  photonic integrated circuits for 6{G} and beyond: A perspective,}
  {\protect\JournalTitle{Journal of Applied Physics}} \textbf{132} (2022).

\bibitem{yin2016realizing}
C.~Yin, Y.~Chen, X.~Jiang, \emph{et~al.}, \enquote{Realizing topological edge
  states in a silicon nitride microring-based photonic integrated circuit,}
  {\protect\JournalTitle{Optics Letters}} \textbf{41}, 4791--4794 (2016).

\bibitem{wu2015scheme}
L.-H. Wu and X.~Hu, \enquote{Scheme for achieving a topological photonic
  crystal by using dielectric material,} {\protect\JournalTitle{Physical Review
  Letters}} \textbf{114}, 223901 (2015).

\bibitem{saby2018}
S.~Barik, A.~Karasahin, C.~Flower, \emph{et~al.}, \enquote{A topological
  quantum optics interface,} {\protect\JournalTitle{Science}} \textbf{359},
  666--668 (2018).

\bibitem{Nik20}
N.~Parappurath, F.~Alpeggiani, L.~Kuipers, and E.~Verhagen, \enquote{Direct
  observation of topological edge states in silicon photonic crystals: Spin,
  dispersion, and chiral routing,} {\protect\JournalTitle{Science Advances}}
  \textbf{6}, eaaw4137 (2020).

\bibitem{lodahl2017chiral}
P.~Lodahl, S.~Mahmoodian, S.~Stobbe, \emph{et~al.}, \enquote{Chiral quantum
  optics,} {\protect\JournalTitle{Nature}} \textbf{541}, 473--480 (2017).

\bibitem{yang2020spin}
Z.-Q. Yang, Z.-K. Shao, H.-Z. Chen, \emph{et~al.},
  \enquote{Spin-momentum-locked edge mode for topological vortex lasing,}
  {\protect\JournalTitle{Physical Review Letters}} \textbf{125}, 013903 (2020).

\end{thebibliography}
\end{document}